\documentstyle[12pt]{article}

\newcommand{\sect}[1]{\setcounter{equation}{0}\section{#1}\indent}

\begin{document}

\topmargin 0pt
\oddsidemargin 5mm
\def\bbox{{\,\lower0.9pt\vbox{\hrule \hbox{\vrule height 0.2 cm
\hskip 0.2 cm \vrule height 0.2 cm}\hrule}\,}}
\newcommand{\AP}[1]{Ann.\ Phys.\ {\bf #1}}
\newcommand{\NP}[1]{Nucl.\ Phys.\ {\bf #1}}
\newcommand{\PL}[1]{Phys.\ Lett.\ {\bf #1}}
\newcommand{\CMP}[1]{Comm.\ Math.\ Phys.\ {\bf #1}}
\newcommand{\PR}[1]{Phys.\ Rev.\ {\bf #1}}
\newcommand{\PRL}[1]{Phys.\ Rev.\ Lett.\ {\bf #1}}
\newcommand{\PTP}[1]{Prog.\ Theor.\ Phys.\ {\bf #1}}
\newcommand{\PTPS}[1]{Prog.\ Theor.\ Phys.\ Suppl.\ {\bf #1}}
\newcommand{\MPL}[1]{Mod.\ Phys.\ Lett.\ {\bf #1}}
\newcommand{\IJMP}[1]{Int.\ Jour.\ Mod.\ Phys.\ {\bf #1}}
\newcommand{\CQG}[1]{Class.\ Quant.\ Grav.\  {\bf #1}}
\newcommand{\PRep}[1]{Phys.\ Rep.\ {\bf #1}}
\newcommand{\RMP}[1]{Rev.\ Mod.\ Phys.{\bf #1}}
\newcommand{\shalf}{\frac{1}{2}}
\newcommand{\pa}{\partial}
\newcommand{\dz}{\frac{dz}{2\pi i}}
\newcommand{\ra}{\rangle}
\newcommand{\lan}{\langle}
\newcommand{\nn}{\nonumber \\}
\newcommand{\vs}[1]{\vspace{#1 mm}}

\def\a{\alpha}
\def\b{\beta}
\def\g{\gamma}
\def\G{\Gamma}
\def\d{\delta}
\def\D{\Delta}
\def\e{\epsilon}
\def\ve{\varepsilon}
\def\z{\zeta}
\def\t{\theta}
\def\vt{\vartheta}
\def\i{\iota}
\def\r{\rho}
\def\vr{\varrho}
\def\k{\kappa}
\def\l{\lambda}
\def\L{\Lambda}
\def\o{\omega}
\def\O{\Omega}
\def\s{\sigma}
\def\S{\Sigma}
\def\vphi{\varphi}
\def\av#1{\langle#1\rangle}
\def\pa{\partial}
\def\na{\nabla}
\def\hg{\hat g}
\def\un{\underline}
\def\ov{\overline}

\begin{titlepage}
\setcounter{page}{0}

\begin{flushright}
COLO-HEP-363 \\
hep-th/9508053 \\
August 1995
\end{flushright}

\vspace{5 mm}
\begin{center}
{\large Remarks on supersymmetric effective actions and supersymmetry
breaking}
\vspace{10 mm}

{\large S. P. de Alwis\footnote{e-mail: dealwis@gopika.colorado.edu}
}\\
{\em Department of Physics, Box 390,
University of Colorado, Boulder, CO 80309}\\
\vspace{5 mm}
\end{center}
\vspace{10 mm}

\centerline{{\bf{Abstract}}}
We discuss some issues related to the definition of different
effective actions,
in connection with the work on supersymmetric theories by Seiberg and
collaborators. We also comment on the possibility of extending this
work to
broken supersymmetric theories.

\end{titlepage}
\newpage
\renewcommand{\thefootnote}{\arabic{footnote}}
\setcounter{footnote}{0}

\setcounter{equation}{0}
\vs{5}
\sect{Introduction}
Recent work by Seiberg and collaborators \cite{seione}  on exact
results in
supersymmetric (SUSY) field theories depends crucially on the
existence of the
so-called Wilson effective action. In contrast to the usual one
particle
irreducible (1PI) effective action, this action, since it has an
explicit
infra-red cutoff, is free from holomorphic anomalies\footnote{This
point seems
to have been highlighted first by Shifman and Vainshtein \cite{sv}} .
This
enables one to make effective use of the power of holomorphy to
constrain the
form of the superpotential that is generated by quantum effects, even
at the
non-perturbative level. As a consequence of this analysis many
remarkable
properties of these field theories (such as the existence of
electro-magnetic
duality in non-abelian theories) have been discovered. It is
therefore of some
importance to discuss the precise (non-perturbative) definition of
the Wilson
effective action in the context of these theories. In this paper we
first
discuss problems associated with defining such an action  in
continuum gauge
theories. Next  (ignoring the above problem) we investigate how such
actions
may be defined for composite fields (mesons, and baryons). In the
fourth
section we discuss questions related to singularities at the origin
of moduli
space. We conclude with an evaluation of the prospects for extending
these
methods to broken SUSY theories.

\sect{Defining a Wilson effective action}
The Wilson effective action is obtained by integrating out short
distance
degrees of freedom of a microscopic theory valid at some short
distance scale,
down to some low energy scale $\L$. It is then an effective field
theory for
discussing physics below the scale $\L$. A precise definition was
originally
given  in the context of
lattice gauge theory \cite{wilson}\footnote{Recently an explicitly
construction
of this has been given \cite{tom}.}.  Unfortunately lattice methods
cannot be
easily extended to supersymmetric theories and one would like to have
a
continuum formulation of the effective action. The first step towards
this was
taken by Polchinski \cite{pol} who derived  a continuum form of the
Wilson
renormalization group equation for scalar $\phi^4$ field theory and
used it to
prove the renormalizability of the theory. The method depended on
explicitly
introducing a
momentum space cutoff in  the kinetic term of the action, and this
enables one
to derive a differential equation for the Wilson action of the form,
\begin{equation}\label{rgeqn}
\L{d\over d\L}S_W (\phi,\L ) = F(S_W,\L)
\end{equation}

The important point about this equation is not so much that it
enables one to
give a proof of  the perturbative renormalization of the theory that
is much
simpler than the usual one, but that it  gives a non-perturbative
definition of
the theory. In other words regardless of the
fact that the derivation of \ref{rgeqn} depended on the existence of
a
perturbative formulation of the theory (in so far as one assumes a
separation
between a kinetic term and an interaction), the equation gives a
unique
non-perturbative definition of an action at any (``final") scale
$\L$, given
the microscopic action at some (``initial") scale $\L_0$. Thus if  we
could
derive such an equation
for SUSY gauge theories we would have a satisfactory object for which
the non-
perturbative arguments of Seiberg and collaborators will apply.

Unfortunately matters are not so simple as soon as one goes beyond
scalar field
theory. The problem arises when one tries to define a regulator for
gauge
theory in order to derive the analog of (\ref{rgeqn}). The
introduction of
higher covariant derivatives by themselves does not help since they
do
not
regularize one loop terms. Thus one is forced to introduce an
additional
regulator (typically Pauli-Villars) to deal with those. Such a
procedure which
regularizes all diagrams (including those in which there are
divergent one-loop
subgraphs) has been given by Warr \cite{warr} and has been extended
to SUSY and
super gravity theories in \cite{kl}. However apart from the rather
baroque
nature of this scheme, it suffers from one serious problem with
regard to
deriving  an equation of the form \ref{rgeqn}. The use of
Pauli-Villars
regulators requires the introduction of preregulators in order to
define each
diagram separately, before the cancellation between diagrams
involving physical
fields and P-V fields takes place. This is done in a fashion that
manifestly
breaks the gauge invariance. This preregulator must  then be first
sent to
infinity. \footnote{It is claimed that at the end of the day one has
a theory
which satisfies the Slavnov-Taylor identities.} The problem is that
the
procedure, even if it yields
a well-defined gauge (or BRST) invariant perturbation series, cannot
be used to
derive the analog of \ref{rgeqn}, since it is essentially a diagram
by diagram
method that is inextricably linked to perturbation theory. It is
thus
important to devise an alternative regularization that enables one to
derive an
RG equation.

In the last few years there has been  quite a bit of activity on this
problem
\cite{wet, bon, becchi, ell}. However none of these attempts seem to
be
completely satisfactory. In the first part of this work we will
derive (in
somewhat simpler fashion and using a different regularization method)
a flow
equation very similar to one  derived in  \cite{wet}, \cite{ell}.
This will
then enable us to elucidate the problems that are involved in
defining a
Wilsonian action.

  A 1PI effective action which is invariant under background field
gauge
transformations  may be defined by the equation \footnote{To keep the
notation
simple we will use the notation of scalar field theory. When relevant
we will
discuss the additional complications of gauge theory such as gauge
fixing and
ghosts. }

\begin{eqnarray}\label{1pi}
e^{-\G (\phi_c)}&=&\int [d\phi ' ]_{\e}e^{-I[\phi_c+\phi ' ]-\phi '.
{\d\G\over\d\phi_c}}\nn
&=&\int [d\phi ' ]_{\e}e^{-[I[\phi _c]+\phi '.K[\phi_c].\phi
'+I_1[\phi_c,\phi
'+J.\phi ' ]}|\nn
&=&e^{-I[\phi _c]}e^{-{1\over 2}tr\ln{K[\phi_c]}-I_1[\phi
_c,-{\d\over\d
J}]}e^{{1\over 4}J.K^{-1}.J}|.
\end{eqnarray}
$I_1$ is at least cubic in the field $\phi'$ and the bar at the end
of the
equation is an instruction to evaluate the expression with
\begin{equation}\label{j}
J=-{\d\G\over\d\phi_c}+{\d I\over\d\phi_c}\equiv\bar J.
\end{equation}
 The subscript $\e$ on the measure is an instruction to cut off the
functional
integral at an ultraviolet cutoff $\L_o^2=\e^{-1}$ and the dot
product implies
integration over space time. In the case of a gauge theory we need to
add
(background covariant) gauge fixing terms and the corresponding FP
term. Since
every term in the functional integral is back ground gauge invariant
so is the
effective action. {\it In addition the BRST invariance of the
functional
integrand and measure implies that $\G [\phi_c]$ is independent of
the gauge
fixing.}

In order to define a Wilson effective action $\G_L[\phi_c]$ we have
to
introduce an infra-red cut off $L =\L^{-2}$. A convenient way of
doing this
while preserving the background gauge invariance, is to modify the
background field propagator  $K[\phi_c ]^{-1}$ using the Schwinger
proper time
representation,  \begin{equation}\label{prop}
K^{-1}_L[\phi_c]=\int_0^L dt e^{-tK[\phi_c]},~~\ln K_L = \int_0^L
{dt\over t}
e^{-tK[\phi_c]}.
\end{equation}
 The Wilson action $\G_L$ is then defined by replacing $K\rightarrow
K_L$ in
(\ref{1pi}). In a gauge theory this would include the contribution
from the
gauge fixing term and so K would explicitly depend on the gauge
fixing
parameter $\xi$ and there would also be a ghost contribution. From
(\ref{1pi})
(with $K\rightarrow K_L$) and (\ref{prop}) we see that
\begin{equation}\label{}
\G_{L\rightarrow\infty} =\G,~~ \G_{L\rightarrow 0} =I[\phi_c],
\end{equation}
(ignoring an infinite constant in the second equation).
To get a flow equation in this background field formalism we need
first to
define an object $W_L[\phi_c, J]$ from equation (\ref{1pi}) by
removing the
restriction $J=\bar J$ (\ref{j}) and of course replacing $K$ by
$K_L$. Then
differentiating this equation with respect to $\ln L$ and
replacing $\phi '\rightarrow  -{\d\over\d J}$ we easily obtain
($\dot{}=L{d\over d L} $)

\begin{equation}\label{w}
\dot{W}_L[\phi_c,J]=tr \dot{K}_L.\left[ {\d W_L\over\d J}{\d
W_L\over\d
J}-{\d\over \d J}{\d W\over\d J} \right].
\end{equation}
Since $K_L$ is linearly divergent as $L\rightarrow 0$ the initial
condition has
to be imposed at $L=\e <<|\phi_c |^{-2}$ and we need to keep the
$O(\e )$ term
so that we have (from the second line
of equation (\ref{1pi}))

\begin{equation}\label{initial}
W_{\e} =I[\phi_c]+{1\over 2}tr\ln K_{\e}[\phi_c]-{1\over
4}J.K^{-1}_{\e}[\phi_c].J
\end{equation}
The equation then gives us $W_L$. Now the Wilsonian action $\G_L$ is
obtained
from this by
putting $J=\bar J$ but since $\bar J$ is defined in terms of $\G_L$
(see
\ref{}) this only gives an implicit definition. To get an explicit
version we
need to Legendre transform with respect to $J$. Putting ${\d
W_L\over\d J}
=<\phi '>_J\equiv \phi_c' $ and
\begin{equation}\label{j}
\bar\G [\phi_c,\phi '_c]=W_L[\phi_c ,J]-J.\phi_c'
\end{equation}
 one immediately has from \ref{w} (using $
 \dot{\bar\G}
=\dot W,~{\d^2W_L\over\d J\d J}=-\left[
{\d\bar\G_L\over\d\phi_c'\phi_c'}
\right]^{-1} $) the flow equation,

\begin{equation}\label{evolution}
\dot{\bar\G}_L[\phi_c,\phi_c']=tr \dot K_L[\phi_c]\left\{
\phi_c'\phi_c'+\left[
{\d\bar\G_L\over\d\phi_c'\phi_c'} \right]^{-1} \right\},
\end{equation}
and the initial condition (\ref{initial}) becomes,

\begin{equation}\label{}
\G_{\e}(\phi_c,\phi_c') = I[\phi_c] +\shalf tr\ln K_{\e}[\phi_c]+
{1\over
4}\phi_c'.K_{\e}.\phi_c'
\end{equation}
In the limit $L\rightarrow\infty$ we have $\phi_c'=<\phi '>_J=<\phi
>-<\phi_c>=0$ and $\bar \G[\phi_c,0 ]=\G[\phi_c]$ the 1PI effective
action.
Hence we may define the Wilson effective action as
\begin{equation}\label{wil}
\G_L[\phi ]= \bar\G_L[\phi_c,0]
\end{equation}
Thus the equation gives us a Wilson effective action at a low scale
$\L
=L^{-1/2}$ given an ``initial" action at a high scale $\L_0
=\e^{-1/2}$.

For a gauge theory the above derivations go through essentially
unchanged
except for the fact
that K now involves terms from the gauge fixing and also from ghosts.
The
effective action obtained this way is background gauge invariant.
However the
introduction of the cutoff $L$ violates the BRST invariance of the
functional
integral.\footnote{I wish to thank Joe Polchinski for  raising this
issue.} The
question is, given the background gauge invariance of the effective
action,
does this matter? Through its dependence on $K$ the right hand side
of the
evolution equation is explicitly dependent on $\xi$ the gauge fixing
parameter.
With generic initial data  we would evolve into an effective action
that is
$\xi$ dependent. On the other hand in the limit $L\rightarrow\infty$
the
functional integral representation implies that BRST invariance is
restored and
the 1PI effective action is
independent of $\xi$. It is hard to understand how this comes about
from the
evolution equation, unless one fine  tunes the initial conditions at
the
physical cutoff  in a $\xi$ dependent manner .  This seems to be
rather an
unsatisfactory state of affairs. Unfortunately  we do not know of any
procedure
for resolving this. Nevertheless we will ignore this gauge fixing
dependence in
the rest of the paper.

\sect{Effective action for composite fields}
As constructed above the Wilson action is not suitable for a
description of the
physics below
the confinement scale in a theory such as QCD, where one expects the
effective
low energy fields to be mesons and baryons. We would need an action
that is
expressed in terms of these
gauge invariant degrees of freedom such as the chiral Lagrangian. So
one may
define (in a schematic though obvious notation), a generating
functional for
mesons in the following way\footnote{The extension to baryons can be
done in an
analogous manner.}.

\begin{eqnarray}\label{wilson1}
e^{iW[J]}&=&\int
e^{i\G_{\e}[Q,\tilde{Q},V]+iJ.Q\tilde{Q}}[dQd\tilde{Q}dV]\int
[dM]\d (M-Q\tilde{Q})\nn
&=&\int [dM] e^{i\bar\G_{\e}+iJ.M}.
\end{eqnarray}
In the above $Q,\tilde Q$ are quark and anti-quark fields,
 $V$ represents the gauge fields and $M$ is a meson field.
Also $G_{\e}$ is the Wilson action valid below the scale
$\e^{-1}$ (so
it should be identified with $\L^2$ of the previous section) and the
functional
integral is effectively cutoff above that scale. An effective action
for mesons
is then
given by,

\begin{equation}\label{}
e^{i\bar\G_{\e}[M]}=\int [dQd\tilde{Q}dV]e^{iS_{\e}[Q,\tilde{Q},V]}\d
(M-Q\tilde{Q}).
\end{equation}
It is important to note that $\bar\G_{\e}[M]$ inherits the
ultra-violet cutoff
so that in calculating meson processes from it one is only supposed
to
integrate up to the scale $\e^{-1}$. Also it is clear that $\bar\G$
has the same
 global symmetry as $S$. It is of course expected that
$M$ will in general acquire a non-zero expectation value and the
action that
describes the degrees of freedom in that vacuum will only exhibit the
diagonal
$ SU(N_f)\times U(1)$ symmetry. It should also be pointed out here
that
$\bar\G$ is a Wilsonian action and is not the same as the 2PI
effective action
\cite{cjt} which is defined by

\begin{equation}\label{}
\G [M_c] =(W[J]-J.M_c)_{M_c={\d W\over\d J}}\simeq
\bar\G_{\e}[M_c]+O(\hbar ),
\end{equation}
the last step following from the saddle point evaluation of the
functional
integral \ref{wilson1}.

In the SUSY case the action $\bar\G$ will  satisfy in particular the
requirements of  holomorphy, so   the arguments of  \cite{seione} as
well as
that of  earlier work \cite{vy},\cite{ads},\cite{amati},  should
apply to this.
In particular this would mean not only that for $N_f>N_c $, $\det
M=0$ but also
that (for $N_f\ge N_c $ the classical constraints (for instance $\det
M =
B\tilde B$) are satisfied. Now while the former result is one that is
used in
\cite{ads} to argue that no superpotential is generated for this
case, the
latter is at variance with the statement in \cite{seitwo}  that the
quantum
moduli space is modified from the classical space, so that for
instance in the
$N_f=N_c$
case the constraint is changed to $\det M- B\tilde{B}=\L^{2N_c}$. In
order to
resolve this  it seems that one has to redefine the Wilson action for
mesons
(and baryons).  To this end let us first write, (again we  explicitly
indicate
only the meson dependence)

\begin{equation}\label{J}
e^{-W[J]}=\int e^{-S_{L}(Q,\tilde{Q}, V)-J.(Q\tilde{Q})}
\end{equation}
The functional integral in the above is taken down to some low
(non-zero )
scale $L^{-\shalf}$
and we define the expectation value in the presence of this cutoff
by,

\begin{equation}\label{exp}
{\d W[J]\over\d J}=<Q\tilde{Q}>_{L ,J}\equiv M_{L, c}
\end{equation}
This expression  agrees with the actual  vacuum expectation value of
the theory
when we let $L\rightarrow\infty$. Now we define the Wilson effective
action for
the mesons by the Legendre transform.

\begin{equation}\label{wilson2}
\G_L(M)=(W[J]-J.M)
\end{equation}
where $J$ is replaced by solving \ref{exp} and we have the equation
of motion
${\d\G\over\d M }=J$.\footnote{\it Henceforth we drop the subscripts
on $M$,
but this object should not be confused with the meson field
introduced earlier,
which was restricted to be $Q\tilde{Q}$.} This is a  Wilsonian action
for
finite $L$. To get the true 1MI (2QI) action one needs to take the
limit
$L\rightarrow\infty$ and we have then a functional of the true vacuum
expectation value\footnote{An interpretation of the effective action
of
\cite{vy}  involving the composite field $S\simeq W^2$ where $W$ is
the gauge
superfield, as a 2PI effective action has  been given earlier in
\cite{fer}.}.
This  will suffer from infra-red problems, but the
Wilsonian
version defined above has an explicit infra-red cutoff and is
therefore
expected to satisfy all the holomorphicity restrictions. The
arguments of
Seiberg et al will therefore  apply to this object. Furthermore we
see from \ref{exp} that quantum effects will prevent us from
concluding that
$\det M=0$ for $3N_c>N_f>N_c$ and hence we may write down a
superpotential

\begin{equation}\label{w}
W={(\det M)^{1\over N_f-N_c}\over \L_s^{3Nc-N_f\over
N_f-N_c}}f(L\L_s^2)
\end{equation}
even in this case ($\L_s$ is the SQCD scale). It should be noted that
this
interpretation of $M$ would appear to resolve the
apparent conflict between \cite{ads} and \cite{vy}.
As pointed out in \cite{seitwo} however there are other problems with
such an
effective action. One is that this superpotential is singular at the
origin\footnote{The superpotential for $N_f<N_c$ is also singular at
the origin
but in that case the potential is minimized away from the origin - at
infinity
in fact, whereas in the present case it is singular - though finite,
at the
point where its minimum lies.}. The other point is that the masssless
degrees
of freedom at the origin of moduli space do not match the quarks and
gluons
of the original formulation.  Equation (\ref{w}) should not therefore
be
considered as an alternative to the discussion of Seiberg for $N_f\ge
N_c$,
nevertheless it is amusing to consider some of its consequences,
especially
since it can be written down once one adopts  (3.4) - (3.6) in
contrast to
(3.2) as the definition of the effective action. We will have more to
say about
the
origin of
moduli space later, but for the time being we
will consider the theory with mass terms for mesons (and baryons) so
that the
minimum is away from the origin. In this case one can still  use the
above (and
its generalizations including baryons) to caculate expressions for
the vacuum
expectation values.  As pointed out in \cite{seione} one can consider
such
terms (for instance a term $m^i_{\hat i}M_i^{\hat i}$) as arising
from
a
coupling to a background field (or external source) and then the
modification
of the field $M=<Q\tilde{Q}>_{L}$ in the presence of this source is
given by
the (analog of) the standard equation for the effective action $\G$
namely
${\d\G\over\d\phi_c}=J$ . Thus  the new expectation value  is
determined by,
\begin{equation}\label{}
{\d W\over\d M^{\tilde{i}}_i} = m_{\tilde{i}}^i
\end{equation}
 with $W$ given by (\ref{w}). This yields (upon taking the limit
$L\rightarrow\infty$ since in the presence of a non-singular mass
matrix $[m]$
there are no infra red singularities ) the result of Amati et al
\cite{amati}
that was used in \cite{seitwo}.

\begin{equation}\label{Mexp}
M^{\tilde i}_{c~i} = <Q_i\tilde{Q}^{\tilde i}>=\L^{3Nc-N_f\over
N_c-N_f}(\det
[m])^{1\over N_c}[m^{-1}]^{\tilde i}_i
\end{equation}
\footnote{In the above we have put  ${f(\infty)\over
Nc-Nf}^{N_c-N_f\over N_c}$
equal to unity. This amounts to a convention for the definition of
$\L$.} As
observed by Seiberg \cite{seitwo} taking  $m\rightarrow 0$ in
different
directions one can generate any value for the meson expectation
values of the
massless theory.

Note that even though (\ref{w}) is not defined for $N_f=N_c$ one can
take this
limit in (\ref{Mexp}) and we get $M=\L^2(\det [m])^{1\over
N_c}[m^{-1}]$ so
that  $\det M =\L^{2N_c}  $

When $N_f>N_c$ we can also define
$B_{L,c}^{f_{N_c+1}...f_{N_f}}=
\e^{f_1...f_{N_f}}<Q_{f_1}...Q_{f_{N_c}}>_L$
where the RHS is defined as in (\ref{J}, \ref{exp}). Then one can for
instance
generate terms of the form $W\sim\L^{\#}B_c^{N_f}\tilde{B}_c^{N_f}$
and  by
adding source terms of the form $bB$ we can  generate non-zero
expectation
values for $B$. The $N_f=N_c$ case is again a limiting one and as
with the
mesons we argue that even though the superpotential at this point is
not
defined one can still first compute the expectation values
(in the presence of the sources) and then take the limit
$N_f\rightarrow N_c$.

The cases $N_f=N_c$ and  $N_f=N_c+1$ were discussed extensively in
\cite{seitwo}. The arguments depend
crucially on the assumption that the constraint on the  quantum
moduli space is
changed (in the $N_f=N_c$ case) from the classical one $\det
M-B\tilde{B}=0$,
to the relation  $\det M-B\tilde{B}=\L_s^{2N_c}$. As pointed out
earlier this
is consistent only with the interpretation of the action given in
(\ref{wilson2}). It should also be noted that
 although with (\ref{wilson1}) we are supposed to integrate down from
the
infra-red cutoff to get the 1MI (and 1BI) actions, with
(\ref{wilson2}) one
simply takes the limit $L\rightarrow\infty$. In other words with the
latter,
one is not supposed to compute loops to
get the 1PI (or 2PI etc.) actions whereas with the former ( the
chiral
lagrangian for QCD for instance), one computes loops.

The most interesting results of the analysis of Seiberg and
collaborators
relate to the
discussion of the theory at the origin of moduli space. This is the
point at
which superpotential is singular and  it is argued that the
description in
terms of $M, B$ becomes invalid for $N_f>N_c+1$. Below we will
discuss some
questions related to the origin of moduli space.

\sect{The origin of moduli space}
Firstly it should be pointed out that since
$M=<Q\tilde{Q}>=\left[{\d\G_{1PI}\over \d Q\d\tilde{Q}}\right]^{-1}$
the point
$M=0$ is a place where the quark inverse propagator blows up.
\footnote{It
should be noted that we are always working with a ultra violet cutoff
so that
products of fields at the same point are well defined.}  It is thus
possible
(as already observed in \cite{seitwo}) that this point is infinitely
far away
in moduli space. Let us examine these  issues more closely.

To begin the discussion let us consider the abelian Higgs model. The
model has
an Abelian gauge field $A_{\mu}$ and a complex scalar field $\phi$
with the
invariance under the $U(1)$ gauge transformations $A_{\mu}\rightarrow
A_{\mu}+\pa_{\mu}\chi $ and $\phi\rightarrow e^{ig\chi}\phi$. The
degrees of
freedom (DOF) in the model are counted as follows: $A_0$ is a
Lagrange
parameter ($\Pi_0=0$) leaving $2\times (3+2)$ phase space dimensions
for the
three spatial gauge fields and two scalar fields. The Gauss law
constraint
${\bf \nabla . E}=j(\phi,\Pi_{\phi})$ and the gauge freedom ${\bf
A\rightarrow
A+\nabla }\chi $ then reduce the
dimension of phase space by two leaving four as the number of  DOF.
This may be
interpreted as two for the (massless) gauge boson and two for the
complex
scalar. Now in this theory it is usually argued that when the
potential for
$\phi$ has a non-trivial minimum
($ V'(\phi_0\ne 0)=0$), the gauge  symmetry is spontaneously broken;
the resulting  Goldstone mode is ``eaten" by the gauge field which
thus
acquires a mass and hence has three degrees of freedom with one more
coming
from real part of the Higgs field. This at least is the classsical
argument.

Quantum mechanically the argument is somewhat more subtle. The
problem is that
in a  quantum gauge theory one can never really ``break" the gauge
invariance.
The reason is that (at least for compact gauge groups) Elitzur's
theorem
\cite{elit} tells us that the
vacuum expectation value of any local order parameter (which does not
have a
gauge singlet component so that its integral over the gauge group
vanishes)
such as $<\phi>$ is zero.\footnote{Strictly speaking this theorem
has been
proved only for a Euclidean lattice gauge theory. In a Minkowski
formulation it
would follow from the Gauss law constraint provided that a
non-perturbative
gauge invariant regularization exists. }  Thus if we construct a
gauge
invariant  (1PI or Wilson)
effective action $\G [A_c,\phi_c]$ the solution of
${\d\G[A_c,\phi_c]\over
\d\phi_c }=0$  is  $<\phi >=0$ which seems to indicate that there is
no
spontaneous symmetry breaking. However it is possible to reformulate
the theory
in terms of gauge invariant variables\footnote{The fact that Higgs
theories can
be expressed in terms of physical gauge invariant variables is an old
story and
was discussed in connection with the absence of a phase transition
between
Higgs and confinement phases in theories with fundamental Higgs in
\cite{banks}. }. One makes the {\it field redefinitions}
\footnote{This
procedure is usually called gauge fixing to the unitary gauge.
However it is
more appropriate to call it a field redefinition. The discussion
later, of the
measure should clarify the difference.} $\phi =e^{ig\theta}\rho
{}~(\rho
>0),~~A_{\mu}=G_{\mu}+\pa_{\mu}\theta$.
Under the gauge transformation $\t\rightarrow \t+\e$; but the fields
$\rho ,
{}~G$ are gauge invariant. Because of  gauge invariance the  $\theta$
dependence drops out and the  classical Lagrangian becomes,
\begin{equation}\label{}
 L=-{1\over 4}(\pa_{\mu}G_{\nu}-\pa_{\nu}G_{\mu})^2-{1\over
2}(\pa_{\mu}\rho
)^2-{1\over 2}g^2\rho^2G_{\mu}^2-V(\rho).
\end{equation}
Obviously when expressed in terms of the physical variables, there is
no gauge
invariance in the Lagrangian. It is instructive to count the degrees
of freedom
(DOF) in this formulation. Again $G_0$ is Lagrange multiplier and so
the phase
space variables are $(\rho ,\Pi_{\rho}),~({\bf G,\Pi =E)}$. These
variables are
unconstrained (the Gauss law  which is now ${\bf\nabla .E}=g^2G_0\rho
$
involves the Lagrange multiplier) and hence the number of degrees of
freedom
are again four, but with one coming from the scalar and three from
the gauge
field. It should be stressed that this has nothing to do with the
shape of the
potential $V$. What  does depend on the latter is the count of
linearized DOF.
Unless $V(\rho )$ has a non-trivial minimum this count will not give
the
correct number of DOF.

What happens in the quantum theory? Here we need to consider the
measure. Under
our change of variables we get  $[dA][d\phi_R][d\phi_I] =[dG][d\t
][\rho d\rho
]$. Since the action is independent of $\t$ the first factor just
gives the
volume of the gauge group. The last factor exhibits the typical
singularity of
polar coordinates and is not in a  suitable form for perturbation
theory. A
further change of variable $\rho\rightarrow\chi =\rho^2$ gives a
linear measure
but now the problem reappears in the kinetic term which becomes
${1\over
8\chi}(\pa\chi)^2$.The metric on ``moduli" space appears to be
singular though
the distance from any point to the origin is in this case is actually
finite
(of course here the singularity is just a coordinate singularity
because the
space is one dimensional). A perturbative evaluation of the
functional integral
requires the existence of a saddle point $\chi_0\ne 0$. Nevertheless
the
expression eqn. (\ref{1pi}) will remain well defined even if (when
the external
source is set to zero) $\chi_c=<\chi >=0$.

These arguments are easily generalized to non-Abelian Higgs theories.
Let us
first consider the case of a Higgs field ${\bf\phi}$ in the
fundamental
representation of the gauge group ${\cal G} = SU(N)$.
Under gauge transformation $\bf\phi\rightarrow g\phi$ and ${\bf
A\rightarrow
gAg^{-1}+gdg^{-1}}$ where ${g\bf }\e {\cal G}$. Again we can redefine
the
fields by putting
$ {\bf \phi =\O\rho }$ (with $\rho =[\rho,0...0]^T$ and
${\bf\O}\e\cal G$ with
$\rho$ a real and positive field) and $\bf A = \O (\pa + G)\O^{-1}$.
The gauge
 transformations are $\O\rightarrow g\O ~\rho\rightarrow\rho,~
G\rightarrow G$,
and the Lagrangian becomes

\begin{equation}\label{}
L=-{1\over 4}tr F^{G~2}-{1\over 2}(D^G\rho )^2-V(\rho ).
\end{equation}
 In these variables the gauge invariance has been reduced from
$SU(N)$ to
$SU(N-1)$. This fact is again independent of the nature of $V$. Let
us now look
at the measure. The original gauge invariant metric on field space is
$\int
d^Dx (\d\phi^{\dag}\d\phi +tr\d A^2)$. Using
$\O^{\dag}\O = \O\O^{\dag}=1$ and the reality of $\rho$, we find that
this
becomes$\int d^Dx(\rho^{\dag}\d\O^{\dag}\d\O\rho +(\d\rho )^2+(\d
G^2))$. Hence
the measure becomes $\prod_{x} [d(G/H)(x)\rho(x)d\rho\prod_{\mu\a}\d
B_{\mu}^a
$, where the first factor is a volume element on $ SU(N)/SU(N-1)$
which will
just give a factor of the volume of the coset in the functional
integral. Again
as in the Abelian case we have the quantum  theory expressed entirely
in terms
of  variables which display a smaller gauge invariance than the
original ones
but as in the earlier case there is no perturbative description
around
$\rho=0$.

By repeating the process when there are $M$ complex scalar fields in
the
fundamental  representation we can express the quantum theory (i.e.
the
classical lagrangian as well as the measure) completely as a SU(N-M)
invariant
theory for $M<N-1$ and as a theory with no gauge invariance for $M\ge
N-1$.
Again this fact has nothing to do with the nature of the classical
potential.

 Consider now a scalar Higgs field in the adjoint representation. We
may do the
field redefinition $\Phi=\O D\O^{-1}$  and $\bf A = \O (\pa +
G)\O^{-1}$, where
$\O\e{\cal G}$ and $D=diag [\l_1,\ldots ,\l_N]$, ($\l_i$ being the
(real)
eigenvalue fields of the Hermitian matrix $\Phi$). As before $\O$
disappears
from the Lagrangian but now we are left with
 a $U(1)^{N-1}$ symmetry (except on a set of measure zero in the
field space
where the symmetry is larger).  The metric on field space may now be
written as
\begin{equation}\label{metadj}
\int d^Dx (tr\d\Phi^2+tr\d A^2)=\int d^Dx \{tr\d D^2+2tr[\O^{-1}\d\O
D
)D-D^2(\O^{-1}\d\O) ]+tr\d G^2\}
\end{equation}
{}From this  we would again expect a  measure of the form $\sim Vol
[SU(N)/U(1)^{N-1}]\prod_x D\d D\d G $, though it is not clear to us
how to
establish this precisely.
Again one expects that because of the singularity at $D=0$ that
perturbation
theory about this point is invalid. Nevertheless the complete
regulated (for example on
a lattice) functional integral expressed as a theory  with  reduced
gauge
invariance is perfectly well defined.

Let us now turn to SUSY theories. Consider first $N=1$ SQCD and for
simplicity
we will take the gauge group to be $SU(2)$ with two doublet chiral
super
fields, $\Phi,~\tilde{\Phi}$. The Lagrangian  takes the form (in
standard
superfield notation) ,
\begin{equation}\label{sqcd}
L={1\over 4\pi}Im\int d^2\theta\tau tr W^{\a}W_{\a}+\int d^4\t
\Phi^{\dag}e^V\Phi +\tilde{\Phi}e^-V^T\tilde{\Phi}+\int d^2\t
W[\Phi,\tilde{\Phi}]+h.c..
\end{equation}
Under a gauge transformation, we have $\Phi\rightarrow e^{i\L}\Phi,
e^V\rightarrow e^{-i\L^{\dag}}e^{V}e^{i\L}$ and the gauge invariant
measure is
obtained from the metric,

\begin{equation}\label{susymet}
||\d V||^2+||\d\Phi ||^2 +...=\int d^xd^4\t tr [(e^{-V}\d e^{V})^2
+\d\Phi^{\dag}e^V\d\Phi+...]
\end{equation}
(the ellipses are for the $\tilde{\Phi}$ terms). This is a highly
non-linear
metric and so is the associated metric. (In SUSY perturbation theory
the
measure may be  replaced by a linear approximation to it). Let us now
try to do
the (analogs of the) field redefinitions that we
performed in non SUSY theories. Thus we put
$\Phi=e^{-i\o}\underline\rho,
\tilde{\Phi}=e^{-i\o}\tilde{\underline\rho}$ where the first factor
in these
expressions is a chiral superfield valued in the gauge group and
$\underline\rho =[\rho,0]^T $. Also we put
$e^U=e^{i\o^{\dag}}e^Ve^{-i\o}$. The
gauge transformation is now $e^{-i\o}\rightarrow e^{-i\L}e^{-i\o}$
but $\o$
disappears from the Lagrangian and we have a classical theory
expressed
entirely in terms of gauge invariant variables. Note that $U$ is no
longer in
Wess Zumino gauge, but (for each generator of the group) the 4
massless degrees
of freedom of the vector multiplet combine with the 4 DOF of the
chiral
multiplet $\o$ to give the 8 DOF of a massive multiplet. This is just
the Higgs
effect in a SUSY theory and again as in the case of the non-SUSY
Higgs theory
discussed above the count of the DOF of the classical non-linear
evolution
equations have nothing to do with the nature of the potential.
Unfortunately in
the SUSY case it is not easy to show that the volume of the gauge
group factors
out. It may be recalled that  in the non-supersymmetric case the
metric became
diagonal because we of the reality of the (gauge invariant) Higgs
field $\rho$.
In the SUSY case this is  necessarily chiral and we cannot make the
same
argument. Nevertheless it is still reasonable to assume that the
qualitative
features of the previous discussion (for example the statement that
perturbation theory around
$\rho =0$ is singular) may still be expected to be valid. In any case
the
potential is given by
$V=G^{i,\bar j}{\pa W\over \pa\phi^i}{\pa \bar{W}\over
\pa\phi^{\bar{j}}}$
where $G$ is the metric on moduli space and it is invariant under
holomorphic
field redefinitions and so there are additional  singularities (i.e.
apart from
the singularities of $W$), only if there is a true singularity of the
metric as
opposed to a coordinate singularity. Our point here is that the
singularity at
the origin coming from the measure appears to be of the latter kind
and hence
should not have any physical effect (in contrast to those of the
superpotential
discussed by Seiberg \cite{seione}).

Next we make a remark about $N=2$ SUSY theories.  In these theories
there is a
$N=1$ chiral superfield in the adjoint representation, and hence as
in the
corresponding case of the non-SUSY theory discussed above,  (modulo
the caveat
about measures in the previous case) the $SU(N)$ theory is equivalent
to a
$U(1)^{N-1}$ theory except that there is no perturbative description
around
$D=0$ (here the D is defined by the SUSY generalization of  the non
SUSY
version given in the paragraph above (\ref{metadj}). As explained in
\cite{sw}
one needs in that case to go to the dual theory, which brings in the
issues of
monopole condensation and the dual Meissner effect. It appears to us
however
that this effect, while it has a gauge invariant formulation in these
$N=2$
theories is not easily taken over as an explanation of confinement in
QCD where
there is no adjoint scalar field to break the symmetry to $U(1)^2$.
The
attempts to gauge fix to the maximal Abelian gauge \cite{'thooft} for
instance
by introducing a term $G^2$ (where $G$ is the gluon field strength)
is
tantamount to introducing a factor $0\over 0$ into the path integral
(since the
integral over the gauge group of any non-trivial irreducible
representation of
the group  is zero)  and is thus bound to yield ambiguous results.

\sect{Supersymmetry breaking}
Finally let us make some remarks on supersymmetry breaking. We will
combine the
old  observations of  Giradello and Grisaru \cite{gg}  on soft
breaking terms,
the recent work of Seiberg and Intrilligator \cite{seione}, and those
of this
paper, to investigate to what extent non-perturbative statements can
be made in
the presence of SUSY breaking.

Consider the SQCD action (\ref{sqcd}) together with the following
terms
coupling  local  gauge invariant composite fields to external
sources.
\begin{eqnarray}\label{break}
S_B &=& \int d^4 x \{
J_TQ^{\dag}e^VQ|_D+J_{\tilde{T}}\tilde{Q}^{\dag}e^{-V^T}\tilde{Q}|_D+
\nn
&+&(J_SW_{\a}W^{\a}+J_MQ\tilde{Q}+J_BQ^{N_c}+
J_{\tilde{B}}\tilde{Q}^{N_c}+h.c.)|_F\}
\end{eqnarray}
 In the above $J_{T(\tilde{T})}$ are real super fields while the
other sources
are chiral superfields. Also for simplicity flavor symmetry
invariance is
preserved.

The computation of correlation functions of gauge invariant composite
fields in
the SUSY theory
are obtained by differentiating the generating functional $W[J]$,
defined in
(\ref{J}), with respect to the sources and then setting them equal to
zero. The
correlation functions of  (softly) broken theory on the other hand
may be
obtained by the same generating functional by doing the functional
differentiations and then setting\footnote{In using the method of
reference
\cite{gg} we find it unnecessary to include a global SUSY breaking
(O'Raffeartaigh) sector as in \cite{evans}. This is just as well
since in
realistic models one expects the soft breaking to be generated from
the
breaking of local SUSY at the Planck scale.}
\begin{equation}\label{susybreak}
J_T=m^2_T\t\t\bar\t\bar\t ,~J_S=m_g\t\t ,~J_M=m_M\t\t,~etc.
\end{equation}

The equations
\begin{eqnarray}\label{}
{\d W\over \d J_T}&=&<Q^{\dag}e^VQ>_J\equiv T,~{\d W\over \d
J_S}=<W_{\a}W^{\a}>_J\equiv S,\nn{\d W\over \d
J_M}&=&<Q\tilde{Q}>\equiv M,
{}~{\d W\over \d J_{B}}=<Q^{N_c}>\equiv B
\end{eqnarray}
 (together with two more for anti-quarks and anti-baryons) define
effective
fields in terms of which an effective action for mesons and baryons,
and
glueballs can be defined. Indeed
if $W$ has been defined in the presence of an infra-red cutoff as in
(\ref{J})
then we can again define a Wilsonian effective action $\G
[T.\tilde{T}, S, M,
B,\tilde{B}]$ as in  (\ref{wilson2}) such that the effective fields (
in the
presence of the source superfields  ) are solutions of  the
equations,
 \begin{equation}\label{}
{\d \G\over \d T}=J_T,~{\d \G\over \d S }=J_S, {\d \G\over \d M}=J_M,
{}~{\d
\G\over \d B}=J_B,~etc..
\end{equation}
The effective action in the presence of SUSY breaking can then be
written as,

\begin{equation}\label{}
\bar\G
[T,...\tilde{B},J]=\G[T,...\tilde{B}]+J_T.T|_D+...+(J_{\tilde{B}}.
\tilde{B}|_F+h.c.)
\end{equation}

with $J$ being replaced by the values given in (\ref{susybreak}).
When the
supersymmetry breaking parameters ($m_T,...m_{\tilde{B}}$) are set to
zero and
$T$ is replaced by using the equation of motion ${\d\G\over\d T}=0 $
giving
$T=aM^{\dag}M+b B^{\dag}B+etc. $,
the
effective action should reduce to those derived in the literature.
However in
the presence of the D-term SUSY breaking  involving the real
superfield $T$
very little can be said about the SUSY broken theory.  This is
because in
solving for T one now has to use  ${\d\G\over\d T}=J_T $ and without
knowing
the dependence  of $\G$
on $T$ it is not possible to extract the dependence of the SUSY
broken action
on the symmetry breaking parameters $M_t,m_{\tilde{T}}$. Of course
one may
consider pure F type symmetry breaking and then one would be able to
calculate
precisely the modifications of the results of Seiberg and
collaborators.
However as is well known such breaking is not phenomenologically
acceptable
since for instance they predict squark (and slepton) masses which are
smaller
than those of   quarks  (and leptons). One can also assume that the
SUSY
breaking parameters are small and then the qualitative features of
the SUSY
limit are preserved \cite{aharony}. Unfortunately in the real world
the SUSY
breaking scale is much larger than the QCD scale. Of course in any
case,
since there is no  chiral scalar superfield description of the proton
for
instance
\footnote{the color antisymmetrization will cause $uud$ to vanish
when each is
a  scalar
super field.}
there is no smooth limit from the effective theory of the mesons and
baryons
defined by Seiberg \cite{seione} \footnote{For instance with
$N_f=N_c$ where
the number
of massless constituent degrees of freedom in the SUSY limit is
matched by the
set of $M$'s
 and $B$'s defined in \cite{seione}. Perhaps this implies that the
proton
composite operator becomes massive.} to an effective description in
terms of
protons, neutrons,  pions etc  of   strongly coupled QCD. A
description in
terms of the observed baryons and mesons, of the   limit  where the
superpartners decouple, which is smoothly obtained from the SUSY
limit, would
seem to require additional operators in the original SUSY theory.
This fact may
have important implications for understanding supersymmetry breaking.

Note added: While preparing this work  for publication we came across
\cite{evans2} in  which aspects of supersymmetry breaking are
discussed.

{\it Acknowledgements}
We are  grateful to R. Leigh and N. Seiberg for very instructive
lectures on
exact results in SUSY field theories and for sharing their insight
into the
subject. We also wish to thank S. Balakrishnan, E. Brezin, M. Dine,
S.
Giddings, A. Hasenfratz, R. Kallosh, D.  MacIntyre, K. Mahanthappa,
M.
Shifman, L. Thorlacius,  and especially, T. DeGrand, J. Polchinski
and L.
Susskind, for discussions. I also wish to thank N. Evans,
S.D.H. Hsu,  M. Schwetz,  S.B. Selipsky, and F. Quevedo for comments
on the
manuscript.
 This work is
partially supported by the
Department of Energy contract No. DE-FG02-91-ER-40672.

\end{document}